\documentstyle[twoside,fleqn,espcrc2]{article}
\include{psfig}

\newcommand{\beqn}{\begin{equation}}
\newcommand{\beqa}{\begin{eqnarray}}
\newcommand{\eeqn}{\end{equation}}
\newcommand{\eeqa}{\end{eqnarray}}
\newcommand{\AmS}{{\protect\the\textfont2
  A\kern-.1667em\lower.5ex\hbox{M}\kern-.125emS}}

\title{Renormalization Group Flow in non-compact QED with two
       charged staggered Fermions}

\author{Arifa Ali Khan\address{HLRZ, c$\setminus$o KFA J\"ulich,
        W-5170 J\"ulich, Germany}}

\begin{document}

\begin{abstract}
We have investigated a system with two sets of staggered fermions
with charges 1 and -1/2
coupling to a non-compact U(1) gauge field in 4 dimensions.
The model exhibits breaking of chiral symmetries of both fermions.
Chiral condensates, renormalized fermion masses and renormalized charges
have been measured.
\end{abstract}

\maketitle

\section{Introduction and Motivation}

Our interest in studying a model with two sets of dynamical fermions
with different charges is motivated by
previous results from non-compact QED with only one charge.
There one finds at strong coupling a chiral symmetry breaking phase
transition of 2nd order at a certain critical coup\-ling
\cite{kogu88,kogu89,boot89,horo90a,schi90,azco92},
which has encouraged speculations
about a possible non-trivial UV-stable fixed point of the Callan-Symanzik
$\beta$ function.
But there is evidence for the vanishing of the
renormalized charge in the continuum limit taken at this critical
coupling \cite{schi92,horo90b}.
Thus one is led to understand QED as a cutoff theory
as is
commonly done with ${\phi}^{4}$-theories, whose trivality is now quite
well established. For a certain value of the renormalized charge
the theory is only valid up to a certain cutoff $\Lambda$ above which new
physics is required. On the other hand, if one sets the cutoff e.g.
equal to the Planck scale, one can obtain, taking
all charged fermions in the standard model into account, a rough
estimate for an upper limit
for the renormalized charge: ${\alpha}_{R} \stackrel{<}{\sim}
1/50$ \cite{schi92}.
But in the standard model, which contains several charged particles
such as electrons, u- and d-quarks, one has to expect a
more complicated phase structure, and a bound obtained from a model with
one charge might have to be
considerably modified. A more precise estimate of the bound would
be important to answer the question whether the
value of the fine structure constant ${\alpha}_{R} = 1/137$
can be explained through the triviality of QED.

Thus by simulating a model with two differently charged fermions we hope
to get more insight into these problems.

\section{The Model and its Phase Structure}

In our simulations we restrict ourselves to a model consisting of two
fermions with a ratio of their charges of -1/2
(in the following called u- and d-quarks),
coupling to a non-compact U(1) gauge field. The gauge field action reads:
\beqn
S_{g} = \frac{\beta}{2} \sum_{x} \sum_{\mu < \nu} F_{\mu \nu}^{2}(x),
\eeqn
\beqn
F_{\mu \nu} (x) = \Delta_{\mu}A_{\nu}(x) - \Delta_{\nu}A_{\mu}(x).
\eeqn
The action for the staggered fermions looks as follows:
\beqa
S_{f} & = & \sum_{x,y}\{ \bar{\chi_{u}}(x)[M_{xy,u}(x)+m_{u}\delta_{xy}]
\chi_{u}(y)  \nonumber \\
& + & \bar{\chi_{d}}(x)[M_{xy,d}(x) +
m_{d}\delta_{xy}]\chi_{d}(y)\},
\eeqa
\beqa
M_{xy,k}(x) & = & \frac{1}{2}\sum_{\mu} \eta_{\mu}(x)
 \{ e^{c_{k}iA_{\mu}(x)}\delta_{y,x+ \hat{\mu}} \nonumber \\ & - &
 e^{-c_{k}iA_{\mu}(y)}\delta_{y,x- \hat{\mu}} \};\nonumber
\eeqa
\beqa
k = u, d & ; & c_{u} = 1, c_{d} = -1/2 .
\eeqa
%

The boundary conditions for the gauge fields are
periodic in all directions and for the fermions
antiperiodic in the time direction and periodic in the spatial
directions.
For the simulations we used a Hybrid Monte Carlo algorithm.
We measure the chiral condensates $\sigma_{i}$ of both types of fermions
at several values of $\beta$ and bare masses $m_{u}$ and $m_{d}$
($m_{i} = $ 0.02, 0.04, 0.09, 0.16).
\begin{figure}[htb]
\vspace{-3.5cm}
\centerline{\psfig{file=ccon1.ps,height=14cm}}
\vspace{-3cm}
\caption{{\em Mean field fit for the chiral condensates}}
\label{fig:chco}
\end{figure}
For each fermion the data were fitted with the equation of
state of a chiral O(2)-symmetric mean field model without coupling
between the two species of
fermions and without logarithmic renormalization
corrections (see fig.~\ref{fig:chco}):
\beqa
m_{i} & = & 2\kappa_{i}\sigma_{i} + 4\zeta_{i}\sigma_{i}^{3}.
\label{eq:mf}
\eeqa
The parameters $\kappa_{i}$ and $\zeta_{i}$ are functions of the coupling
$\beta$:
\beqa
\kappa_{i}/\zeta_{i} & = & \bar{\kappa}_{i}(1-\beta/
\beta_{ci}), \nonumber \\
1/\zeta_{i} & = & \bar{\zeta}_{i}+\bar{\bar{\zeta}}_{i}
(1 - \beta/\beta_{ci}) .
\eeqa
The critical couplings for the chiral transitions in the mean field model
are denoted by $\beta_{ci}$.
We find $\beta_{cu} = 0.183(1)$ and $\beta_{cd} = 0.049(1)$.
The fitted equations of state give a quite good description for the
$\sigma_{u}$ data for $0.15 \leq \beta \leq 0.22$
and for the $\sigma_{d}$ data for $0.04 \leq \beta \leq 0.22$.
Thus the two fermions appear to have chiral transitions
at different values of the bare coupling.
This can be considered as an indication against confinement driven by
monopole condensation (monopole condensation is claimed to occur in
non-compact lattice QED by Hands et al. \cite{hand91}), because
if chiral symmetry breaking was coupled to confinement
in this theory, as soon as one particle condenses there should be no
free charges at all in the spectrum.

\section{Renormalized Fermion Masses}
Because the fermions are charged the fermion correlation functions
are gauge dependent. To measure the fermion masses we have to fix the
gauge. We use Landau gauge, but there are still certain gauge degrees of
freedom which remain
unfixed. To fix them, we restrict the lattice average
$\bar{A_{\mu}}$ of the
gauge field between $-\pi/L_{\mu}$ and $\pi/L_{\mu}$.
Because $\bar{A_{\mu}} \neq 0$ we have to fit the correlation
functions
with free fermion propagators
in a constant background field given by the average of $\bar{A_{\mu}}$
over a set of configurations.

The masses of both fermions go to zero as one approaches $\beta_{cu}$,
but near $\beta_{cd}$ the u masses are O(1) in lattice units.
A continuum limit with both particles in the spectrum can therefore
be performed only at $\beta_{cu}$.

\section{Renormalized Charges}
Due to the Ward identity the renormalized charge can be related to the
bare charge through the photon wave function renormalization:
\beqn
e_{R\,u,d}^{2} = Z_{3}\: e_{u,d}^{2}.
\eeqn
$Z_{3}$ can be obtained by extrapolating the gauge invariant part of the
vacuum polarization function to momentum zero:
\beqn
Z_{3} = \lim_{k \rightarrow 0} D(k) .
\eeqn
Because the photon-photon correlator itself shows very large
fluctuations, we calculated $D(k)$ from a correlator between the fermion
current and the photon field:
\beqn
D(k) = 1 - \left.\frac{1}{N_{k}V} \sum <\tilde{j_{\mu}}(k)
\tilde{A_{\mu}}^{*}(k)> \right|_{k_{\mu}=0}.
\eeqn
The sum runs over all directions $\mu$ and over all
$k$ with $k^{2}$ fixed and $k_{\mu}=0$, and
$N_{k}$ is the number of possibilities to choose such a $k$.
The resulting renormalized $\beta$ can be compared with the
result obtained from one-loop lattice perturbation
theory with the corresponding bare coupling and the renormalized masses
as input:
\beqn
\beta_{R} - \beta = -\Pi(k=0,m_{uR},m_{dR},V=\infty) \label{eq:pt}
\eeqn
As also in QED with only one charge,
we find good agreement between the data and the
perturbative results (see figs.~\ref{fig:rech1} and ~\ref{fig:rech2}).

\begin{figure}[t]
\vspace{-3.5cm}
\centerline{\psfig{file=brmr1.ps,height=12cm}}
\vspace{-1cm}
\caption{{\em Renormalized coupling as a function of renormalized
u and d masses}}
\label{fig:rech1}
\end{figure}
\begin{figure}[t]
\vspace{-3.5cm}
\centerline{\psfig{file=brmr2.ps,height=12cm}}
\vspace{-1cm}
\caption{{\em same as fig. 2, sideview}}
\label{fig:rech2}
\end{figure}

\section{How to find the maximal Cutoff?}

We have seen that the dependence of $\beta_{R}$ on the renormalized
masses can be well described by renormalized perturbation theory.
This reassures us about the triviality of this model
and gives us a tool to compute triviality bounds for a given $\beta_{R}$.
For small masses the perturbative formula (\ref{eq:pt}) yields
a linear behaviour of $\beta_{R}$ with the logarithms of the masses:
\beqn
\beta_{R} - \beta = -\frac{2}{3\pi^{2}}\ln m_{uR} - \frac{1}{6\pi^{2}}
\ln m_{dR} + a.
\label{eq:sm}
\eeqn
$a \simeq 5/4\cdot 0.0210$ \cite{schi92}.
If we now want to change the cutoff while keeping the renormalized
charge fixed we must take care to stay at a fixed ratio
$R = m_{uR}/m_{dR}$ of the renormalized masses. From (\ref{eq:sm}) we
obtain
\beqn
\beta_{R} - \beta -\frac{2}{3\pi^{2}}\ln R =
- \frac{5}{6\pi^{2}} \ln m_{uR} + a .
\label{eq:cm}
\eeqn
We now have to describe $m_{iR}$ as functions of the bare quantities.
Assuming that it should be possible to describe the renormalized masses
with a gap equation which relates the chiral
condensate to the trace of the renormalized fermion pro\-pagator and
additional terms linear in the bare masses,
we fitted  $m_{iR}$
as 3rd order polynomials in $\sigma_{i}$ with coefficients
dependent on $(1 - \beta/\beta_{ci})$ and linear in $m_{u}$ and $m_{d}$.
With the help of the mean field equations of state (\ref{eq:mf})
for the chiral condensates in the parameter
region around $\beta_{cu}$
and the one-loop formula for $\beta_{R}$ we calculated $\beta_{R}$
on a grid of points $0.15 \leq \beta_{\rm i} \leq 0.223$, i = 1, 12;
$0.02 \leq m_{d {\rm j}} \leq 0.12$, j = 1, 10; and
$0.02 \leq m_{u {\rm k}} \leq 0.12$, k = 1, 19. With an
interpolation linear in $\log m_{u}$, $\log m_{d}$ and linear in $\beta$
we can produce surfaces $\beta_{R} = const$ and find lines
with constant mass ratios on them.

In fig.~\ref{fig:betc} we have plotted a
surface for a small $\beta_{R}$ value and intersection
lines for two mass ratios.
The shape of the lines is dependent
on $R$. For small $\beta_{R}$ the end line of the surface on the
$m_{u} = 0$ plane appears to lie at $\beta$ values close to $\beta_{cu}$,
except for very small $m_{d}$. Because on the intersection lines
the momentum cutoff continuously rises as one lowers $\beta$
in the parameter range considered here,
one expects the maximal cutoff to be reached at a
point $(m_{d0},\beta_{0},m_{u0}=0)$ with
$\beta_{0} \simeq \beta_{cu}$.
In this case we can
use $\beta_{cu}$ as lower bound on $\beta$ in (\ref{eq:cm}) and obtain
a lower limit on $m_{uR}$:
\beqn
m_{uR} \stackrel{>}{\sim} 1.365 \: R^{\frac{4}{5}}
\exp{-\left[\frac{6\pi^{2}}{5}(\beta_{R}-\beta_{cu})\right]}.
\label{eq:mlim}
\eeqn
Even for $R = 1$ and small $\beta_{R}$ the bound is
somewhat lower than the limit obtained from
the one charge model with the same $\beta_{R}$.
This is due to the effects of the d quark, which
causes a
lowering of the critical point $\beta_{cu}$ compared to $\beta_{c}$
used in the analogous formula in the one charge model
(see \cite{schi92}).
However, for mass ratios much different from 1 the result
changes due to the R-dependency in (\ref{eq:mlim}) and because
$\beta_{cu}$ will have to be substituted by a $\beta_{0}$ which can be
much smaller than $\beta_{cu}$.
We want to extend our calculation of $\beta_{R}$ as a function of
the bare quantities also to smaller
bare masses and $\beta$ values in the whole range down to $\beta_{cd}$
to obtain more precise informations about the lines of constant physics.

Another consequence
of the relation between the renormalized masses and $\beta_{R}$
which we have now obtained is that
we can see (e.g. after exponentiation of (\ref{eq:cm}))
that in this theory one cannot let
only one of the renormalized quark masses be zero except
in the limiting case $\beta_{R} = \infty$.
\begin{figure}[htb]
\vspace{-3.5cm}
\centerline{\psfig{file=bconst1.ps,height=12cm}}
\vspace{-1cm}
\caption{{\em Surface $\beta_{R}$ = 0.32 with intersection lines
belonging to mass ratios $R \simeq 6$ (left) and $R \simeq 2.2$ (right).
The black triangle denotes the
location of the critical point $\beta_{cu}$.}}
\label{fig:betc}
\end{figure}


\section*{Acknowledgements}

I am indebted to G. Schierholz, R. Horsley, M. G\"ockeler and
P. Rakow for many discussions and for providing their computer
programs as a basis for my programming.
The numerical computations have been performed on the Cray Y-MP
of the HLRZ J\"ulich and the Alliant at the GMD in Sankt Augustin.
I thank
both institutions for granting the necessary computer time.

\end{document}